\documentclass[prc,preprint,showpacs]{revtex4}
\usepackage{graphicx,dcolumn,array,bm,amsmath,amssymb}

\newcommand{\nc}{\newcommand}               %
\nc{\nuc}[2]	{$^{#1}${#2}}		    
\nc{\hatf}[3]   {\hat{{#1}_{#2}}(#3)}       
\nc{\hatfpm}[4] {\hat{{#1}}_{#2}^{(#4)}(#3)}
\nc{\fcpm}[4]   {{#1}_{#2}^{C(#4)}(#3)}     
\nc{\csop}      {U(\theta)}                 
\nc{\csopd}     {U^{\dagger}(\theta)}       
\nc{\bra}	{\langle}		    
\nc{\ket}	{\rangle}		    
\nc{\br}	{{\bf r}}		    %
\nc{\bp}	{{\bf p}}		    %
\nc{\bk}	{{\bf k}}		    %
\nc{\bhr}	{{\bf{\hat r}}}		    %
\nc{\bhp}	{{\bf{\hat p}}}		    %
\nc{\bR}	{{\bf R}}		    %
\nc{\lam}	{\lambda}		    %
\nc{\al}	{\alpha}		    %
\nc{\gam}	{\gamma}	            %
\nc{\vc}[1]	{\bf #1}	            %
\nc{\Nc}	{M}	                    %
\nc{\NcB}	{M^{B}}	                    %

\begin{document}

\title{Two-body Coulomb scattering and complex scaling}

\author{I. Hornyak}
\email{ihornyak@atomki.hu}
\affiliation{University of Debrecen, Faculty of Informatics, PO Box 12, 
4010 Debrecen, Hungary}

\author{A.T. Kruppa}
\email{atk@atomki.hu}
\affiliation{Institute of Nuclear Research, Bem t\'er 18/c, 
4026 Debrecen, Hungary}

\date{\today}

\begin{abstract}

The two-body Coulomb scattering problem is solved using the standard complex scaling method.
The explicit enforcement of the scattering boundary condition is avoided. Splitting of the 
scattering wave function based on the Coulomb modified plane wave is considered. 
This decomposition leads a  
three-dimensional Schr\"odinger equation with source term. 
Partial wave expansion is carried out and 
the asymptotic form of the solution is determined.
This splitting does not lead to
simplification of the scattering boundary condition if complex scaling is invoked.
A new splitting carried out only on partial wave level is introduced and this method is  proved to be very useful. 
The scattered part of the wave function 
tends to zero at large 
inter-particle distance. 
This property permits of easy numerical solution: 
the scattered part of the wave function can be
expanded on bound-state type basis. The new method can be applied not only for pure Coulomb potential but
in the presence of short range interaction too. 
\end{abstract}


\maketitle

\section{Introduction}

The method of complex scaling (CS) has been an excellent tool
to calculate half life times of resonance states for a long time.
The CS has been successfully  applied in many areas of quantum physics \cite{ho83,moi98} and 
it has been extended to collision processes very early
on \cite{nut69,hen72}. However, a drawback of 
the standard CS (or the uniform CS) has emerged immediately after the introduction of 
the method. For scattering problems the  CS 
procedure can be applied only for 
short range potentials \cite{nut69,bau75}. This is 
indeed serious since the long range Coulomb interaction can not be neglected in
majority of the problems of 
atomic and nuclear physics. Several modifications have been suggested \cite {joh84,res85,pes92} 
but none of them has reached a widespread acceptance. 
After these initial applications the scattering aspects of the CS
has been neglected. 

The turning point has been the work \cite{res97} where
it has been shown that scattering calculations with the exterior CS can be successfully
performed for long range interactions. After this pioneering work the exterior CS method 
has been applied 
for variety of three-body Coulomb problems, even above the three-body breakup threshold,
with great success \cite{bar06,cur04a,cur04b,bae01}. The exterior CS method has proved to be one of the most
successful numerical methods to deal with collision processes. 
However, recently the exterior CS method has been under scrutiny since in the method 
an artificial cutoff in some of the
interaction is used. To solve this problem a modification of the original exterior CS 
method has been suggested and checked in two-body calculations \cite{Vol09,Yak10}. Extension to 
three-body problem has been also sketched \cite{Ela09}.  

Recently it has been shown \cite{kru07} that the standard CS can be applied for scattering problems
when a short range potential is
added to the pure Coulomb interaction. The method is based on the two potential formalism.
Similar approach has been suggested also in \cite{cur04}. In the present paper we 
rigorously develop a method which is 
equally good 
for pure Coulomb interaction and for the general case too 
(i.e. a short range potential is added  to the Coulomb interaction).
The new approach does not rely on the two potential formalism and dangerous cutoff will not be introduced. 

In the case of a two-body problem the wave
function depends on the inter-particle coordinate $\br$. 
The scattering solution of the Schr\"odinger 
equation with momentum $\bk$ is denoted by $\psi^+(\bk,\br)$. This wave function will be called 
three dimensional (3D) wave
function. It is assumed that the wave
function satisfies appropriate scattering boundary conditions. The aim of the application of 
any CS method
is to introduce a new equation instead of the Schr\"odinger equation with simplified boundary conditions. 
The expectation is that 
the solution of this new equation  
is square integrable therefore it can be approximated by bound-state type basis functions. 
In this way the explicit use
of the complicated scattering  asymptotic form of the wave function can be avoided and the numerical 
calculation can be simplified.

In contrast to resonance state calculation in scattering problem the CS is not applied  
directly to the full wave function.  First a splitting of the total wave function is carried out. 
The full scattering solution is searched in the form 
\begin{equation}\label{3dsplit}
\psi^+(\bk,\br)=\phi_0(\bk,\br)+\psi^{sc+}(\bk,\br),
\end{equation}
where  $\phi_0(\bk,\br)$ is a known function. 
From the Schr\"odinger equation for the scattered part of the wave function the so called driven
Schr\"odinger equation (or Schr\"odinger equation with source)
\begin{equation}\label{drsch}
(E-\hat H)\psi^{sc+}(\bk,\br)=S(\bk,\br),
\end{equation}
can be derived. The source term is given by 
$
S(\bk,\br)=(\hat H-E)\phi_0(\bk,\br).
$
The Hamiltonian and energy are denoted by $\hat H$ and $E$ respectively. 

We mention that the two-body Coulomb problem with source has been recently 
thoroughly investigated in \cite{Gas10}. Complicated but exact solutions have been given for 
very general sources. 
Basis functions with proper two-body scattering asymptotic have been generated from the exact solutions 
and used in the J-matrix method \cite{Anc11}.
The driven Schr\"odinger equation has been applied for realistic three-body 
scattering problems too \cite{ran11,fra10}. However, in these works CS has not been applied and the complicated scattering
boundary conditions have been implemented using either the finite element method or Sturmian expansion.

The CS in scattering calculations means that the coordinate $\br$ in Eq. (\ref{drsch}) is 
replaced by $\br e^{i\theta}$ where $0<\theta<\pi/2$. 
The boundary condition is simplified if after 
complex scaling the scattered part of the wave
function goes to zero when the inter-particle distance tends to infinity. It is easy too see that this property 
is fulfilled if 
$\psi^{sc+}(\bk,\br)$ contains only outgoing spherical wave. In the paper this property
will be investigated for different splittings of the full wave function.

The organization of the paper is the following. In section \ref{exact} the known expressions of the
two-body  Coulomb scattering are reviewed. The driven Schr\"odinger equation is introduced in section \ref{driveneq}.
The splitting of the total wave function in Eq. (\ref{3dsplit}) is carried out on 3D level, however, 
the splitting can be carried out only on 
partial wave (p.w.) level.
Different 3D and p.w. splittings will be considered and it will be shown that there are cases when the 3D and p.w. splittings
are not equivalent. The Coulomb modified plane wave (CMPW) plays a basic role in the recent surface
integral formalism of
the scattering theory \cite{Kad09,Kad05}. The properties of the 3D splitting based on the CMPW  
will be investigated in section \ref{cmpw}.
A useful p.w. splitting from the point of view of the CS will be introduced in section \ref{cs}. Finally 
numerical examples will be presented both for the pure Coulomb case and for a potential having short and long range
parts. The conclusions will be given in section \ref{sum}.  

\section{Exact solutions of the two-body Coulomb problem}\label{exact}
First we collect a few known expressions \cite{har} for the two-body Coulomb scattering in order to 
fix the notations.
As usual we take $\hbar=m=e=1$ (m is the reduced mass), the energy is $E=k^2/2>0$ and the Coulomb-potential reads
$\gamma k/r$, where $\gamma$ is the Sommerfeld-parameter. We consider the Schr\"odinger equation with pure two-body
Coulomb interaction
\begin{equation}\label{cousch}
\left(-\frac{1}{2}\Delta_{\br}+\frac{\gamma k}{r}-\frac{k^2}{2}\right) \psi(\bk,\br)=0,
\end{equation}
where $\Delta_{\br}$ is the Laplace-operator. The Coulomb
scattering state 
\begin{equation}
\psi_c^+(\bk,\br)=e^{-\pi\gamma/2}\Gamma(1+i\gamma)e^{i\bk\br}M(-i\gamma,1,ikr-i\bk\br)
\end{equation} 
is a solution of (\ref{cousch}). Here $M(a,b,z)$ is the regular confluent hypergeometric function \cite{abra}.
The partial wave expansion is given by the well known form
\begin{equation}\label{pwcou}
\psi_c^+(\bk,\br)=\sum_{l=0}^\infty (2l+1)\psi^+_l(k,r)P_l(\cos\vartheta),
\end{equation}
where $P_l(z)$ is the Legendre polynomial and $\vartheta$ is the angle between the vectors $\bk$ and $\br$. 
The full radial part $\psi^+_l(k,r)$ is expressed with the help of 
the regular Coulomb function $F_l(k,r)$
\begin{equation}\label{fullcou}
\psi^+_l(k,r)=\frac{1}{kr}i^l \exp(i\sigma_l)F_l(k,r).
\end{equation} 
The explicit formula reads
\begin{equation}\label{coureg}
\psi^+_l(k,r)=\frac{\Gamma(l+1+i\gamma)}{\Gamma(2l+2)}e^{-\gamma\pi/2}e^{-ikr}(2ikr)^l M(l+1-i\gamma,2l+2,2ikr)
\end{equation}
and the Coulomb phase shift is defined by
$e^{2i\sigma_l}=\Gamma(l+1+i\gamma)/\Gamma(l+1-i\gamma)$.
The p.w. components $\psi^+_l(k,r)$ satisfy the radial Schr\"odinger equation 
\begin{equation}\label{radsch}
\left[-\frac{1}{2r}\frac{d^2 }{dr^2}r
+\frac{l(l+1)}{2r^2}+\frac{\gamma k}{r}-\frac{k^2}{2}\right]\psi_l(k,r)=0.
\end{equation}

The Coulomb scattering function can be split into so called incoming and scattered waves \cite{har2}.
Using the identity 7.2.2.9 in \cite{prud} we can write
\begin{equation}
\psi_c^+(\bk,\br)=\psi_i(\bk,\br)+\psi_s(\bk,\br),
\end{equation}
where
\begin{equation}\label{coui3d}
\psi_i(\bk,\br)=e^{\pi\gamma/2}e^{i\bk\br}U(-i\gamma,1,ikr-i\bk\br),
\end{equation}
and
\begin{equation}\label{cous3d}
\psi_s(\bk,\br)=e^{\pi\gamma/2}\frac{\Gamma(1+i\gamma)}{\Gamma(-i\gamma)}e^{ikr}U(1+i\gamma,1,i\bk\br-ikr).
\end{equation}
The notation 
$U(a,b,z)$ stands for the irregular confluent hypergeometric function \cite{abra}.
Interestingly not only $\psi_c^+(\bk,\br)$ but the functions $\psi_{i}(\bk,\br)$ and $\psi_{s}(\bk,\br)$ 
satisfy the 3D Scr\"odinger equation (\ref{cousch}).

The partial wave expansions of the incoming and scattered parts are given in \cite{har2}. Later we will use them so we 
quote the main result of paper \cite{har2}. We use a very similar notation as in \cite{har2} however we have
rewritten the  Whittaker function $W$ 
in terms of $U$. 

The p.w. expansions of $\psi_i(\bk,\br)$ and $\psi_s(\bk,\br)$ are given in the same form as 
(\ref{pwcou}) but the p.w. components now read
\begin{equation}\label{coui}
\psi_{i,l}(k,r)=\omega_{i,l}(k,r)+\chi_l(k,r)
\end{equation}
and
\begin{equation}\label{cous}
\psi_{s,l}(k,r)=\omega_{s,l}(k,r)-\chi_l(k,r).
\end{equation}
We note that our definitions of $\omega_{i,l}$, $\omega_{s,l}$ and $\chi_l$ are constant times of the original ones \cite{har2}.
The explicit expressions are the followings
\begin{equation}\label{pwcoui}
\omega_{i,l}(k,r)=e^{-ikr}e^{\gamma\pi/2}(-1)^{l+1}(2ikr)^lU(l+1-i\gamma,2l+2,2ikr),
\end{equation}
\begin{equation}\label{pwcous}
\omega_{s,l}(k,r)=e^{ikr}e^{2i\sigma_l+\gamma\pi/2}(-1)^{l+1}(2ikr)^lU(l+1+i\gamma,2l+2,-2ikr)
\end{equation}
and
\begin{equation}\label{pwchi}
\chi_l(k,r)=\frac{e^{ikr+\gamma\pi/2}}{2ikr}\frac{(-1)^l}{(2ikr)^l}\frac{\Gamma(2l+1)}{\Gamma(l+1-i\gamma)}\sum_{n=0}^l
\frac{(-1)^n(i\gamma-l)_n}{(-2l)_nn!}(2ikr)^n.
\end{equation}
The equation
\begin{equation}\label{pwcousplit}
\psi^+_l(k,r)=\omega_{i,l}(k,r)+\omega_{s,l}(k,r)
\end{equation}
is also proved in \cite{har2}. We mention that the splitting (\ref{pwcousplit}) simply follows from (\ref{coureg}) if the 
function $M(l+1-i\gamma,2l+2,2ikr)$ is rewritten in terms of the irregular 
confluent hypergeometric functions U using the equation 7.2.2.9 of \cite{prud}.

Although the functions $\psi_i(\bk,\br)$ and $\psi_s(\bk,\br)$ are solutions of the 3D Schr\"odinger equation (\ref{cousch}) 
surprisingly its partial wave components $\psi_{i,l}(k,r)$ and $\psi_{s,l}(k,r)$ do not satisfy the radial Schr\"odinger equation (\ref{radsch}) 
(for details see \cite{har2}). This property will be proved to be very important for our new method.

Using the asymptotic expansion 13.5.2 of \cite{abra} we get the asymptotic form of 
the Coulomb-scattering wave function in the well known form
\begin{eqnarray}\label{casy}
\psi_c^+(\bk,\br)=e^{i\bk\br}(kr -\bk\br)^{i\gamma}\left[1+O\left(\frac{1}{kr}\right)\right]+
f_c(\cos(\vartheta))\frac{e^{ikr-i\gamma\ln(2kr)}}{r}\left[1+O\left(\frac{1}{kr}\right)\right],
\end{eqnarray}
where $f_c(\cos(\vartheta))$ is the Coulomb scattering amplitude. The function $e^{i\bk\br}(kr -\bk\br)^{i\gamma}$ is called
CMPW.

\section{Driven Scr\"odinger equation}\label{driveneq}

The scattering solution of the Scr\"odinger equation is searched in the form (\ref{3dsplit}). 
From the Scr\"odinger equation (\ref{cousch}) with a simple rearrangement  the following driven 
Schr\"odinger equation (or Scr\"odinger equation with source) 
\begin{equation}\label{drivensch}
\left(\frac{p^2}{2}+\frac{1}{2}\Delta_{\br}-\frac{\gamma k}{r}\right)\psi^{sc+}(\bk,\br)=S(\bk,\br),
\end{equation}
can be derived for  $\psi_c^{sc+}(\bk,\br)$.
The source term is given by
\begin{equation}\label{sdef}
S(\bk,\br)=\left(-\frac{1}{2}\Delta_{\br}+\frac{\gamma k}{r}-\frac{k^2}{2}\right)\phi_0(\bk,\br).
\end{equation}
We mention that the driven Scr\"odinger equation (\ref{drivensch}) has been studied in \cite{Gas10}. For quite 
general sources very complicated 
analytic solutions can be found \cite{Gas10}. 
The aim of our paper is to derive an easy numerical method to solve (\ref{drivensch}) and 
from the scattered part of the wave function deduce 
the scattering amplitude. 

The asymptotic form (\ref{casy}) inspires 
the following choice for $\phi_0(\bk,\br)$ 
\begin{equation}\label{cmpwsplit}
\phi_0(\bk,\br)=e^{i\bk\br}(kr -\bk\br)^{i\gamma}.
\end{equation}
This splitting is based on the CMPW and it has been used in \cite{Kad09,Kad05} in order to derive the surface integral formalism of the 
scattering theory. 
Using Descartes-coordinates 
it is easy to derive a simple form for the source term
\begin{equation}\label{sdef11}
S(\bk,\br)=\frac{\gamma^2 k}{r(kr-\bk\br)}e^{i\bk\br}(kr -\bk\br)^{i\gamma}.
\end{equation}

We may try to use the splitting based in the incoming Coulomb wave function  i.e. we make 
the following choice
\begin{equation}\label{harsplit}
\phi_0(\bk,\br)=\psi_{i}(\bk,\br)
\end{equation}
instead of (\ref{cmpwsplit}). In this case we get $S(\bk,\br)=0$. 
This follows from the fact the function $\psi_{i}(\bk,\br)$ satisfies  
(\ref{cousch}). 
In this case we do not get a driven Schr\"odinger equation, $\psi^{sc+}(\bk,\br)$ satisfies the 
original homogeneous equation (\ref{cousch}) and
$\psi^{sc+}(\bk,\br)=\psi_s(\bk,\br)$.

If we want to derive the p.w. form of the driven Scr\"odinger equation (\ref{drivensch}) 
we have to have the p.w. expansions
of the source term 
\begin{equation}\label{sourcepw0}
S(\bk,\br)=\sum_{l=0}^\infty (2l+1)S_l(k,r)P_l(\cos(\vartheta))
\end{equation}
and of $\phi_0(\bk,\br)$
\begin{equation}\label{pwexp2}
\phi_0(\bk,\br)=\sum_{l=0}^\infty (2l+1)\phi_{0,l}(k,r)P_l(\cos(\vartheta)).
\end{equation}
Using the operator identity \cite{mes}
\begin{equation}\label{lapform}
\Delta_{\br}=\frac{1}{r}\frac{d^2}{dr^2}r-\frac{\hat L^2}{r^2},
\end{equation}
where $\hat L^2$ is the square of the orbital angular momentum operator, we can derive 
the partial wave form of the driven Schr\"odinger equation (\ref{drivensch}) 
\begin{equation}\label{pwdriven}
\left[\frac{k^2}{2}+\frac{1}{2r}\frac{d^2 }{dr^2}r
-\frac{l(l+1)}{2r^2}-\frac{\gamma k}{r}\right]\psi_l^{sc+}(k,r)=S_l(k,r),
\end{equation}
where
\begin{equation}\label{pwexp1}
\psi^{sc+}(\bk,\br)=\sum_{l=0}^\infty (2l+1)\psi^{sc+}_l(k,r)P_l(\cos(\vartheta)).
\end{equation}

Later it will be proved to be very useful if we make the splitting of 
the scattering wave function not in the 3D form (\ref{3dsplit}) but on the p.w. level.
We take the p.w. component of the scattering wave function in the following form
\begin{equation}\label{pwsum}
\psi_l^+(k,r)=\tilde\phi_{0,l}(k,r)+\tilde\psi_l^{sc+}(k,r),
\end{equation}
where $\tilde\phi_{0,l}(k,r)$ is a fixed known function and $\tilde\psi_l^{sc+}(k,r)$ is considered as an unknown function.
From the partial wave Schr\"odinger equation we get the following non-homogeneous differential equation
\begin{equation}\label{pwdriven1}
\left[\frac{k^2}{2}+\frac{1}{2r}\frac{d^2 }{dr^2}r
-\frac{l(l+1)}{2r^2}-\frac{\gamma k}{r}\right]\tilde\psi_l^{sc+}(k,r)=\tilde S_l(k,r)
\end{equation}  
where
\begin{equation}\label{pwsterm}
\tilde S_l(k,r)=\left[-\frac{1}{2r}\frac{d^2 }{dr^2}r
+\frac{l(l+1)}{2r^2}+\frac{\gamma k}{r}-\frac{k^2}{2}\right]\tilde\phi_{0,l}(k,r).
\end{equation}  

If we take $\tilde\phi_{0,l}(k,r)$ identical to the partial wave component of $\phi_0(\bk,\br)$ i.e. 
$\tilde\phi_{0,l}(k,r)=\phi_{0,l}(k,r)$ then 
the source terms $S_l(k,r)$ and $\tilde S_l(k,r)$ are identical 
if in equation (\ref{sdef}) 
the action of the Laplace-operator can be given in the form (\ref{lapform}).
This replacement however 
is valid only for those functions which are finite at $r=0$ (see page 496 \cite{mes}).

In the case of splitting based on (\ref{harsplit}) the function $\psi_{i}(\bk,\br)$ is not finite at $r=0$.
In this circumstance the 3D splitting and the p.w. level splitting are different. 
We have already seen that the 3D splitting based on (\ref{harsplit})
does not lead to a driven Schr\"odinger equation. However if we make the following p.w. splitting 
\begin{equation}\label{pwsumhar}
\psi_l^+(k,r)=\psi_{i,l}(k,r)+\tilde\psi_l^{sc+}(k,r),
\end{equation}
i.e we take 
$\tilde\phi_{0,l}(k,r)=\psi_{i,l}(k,r)$ then we get a driven radial Schr\"odinger equation.
Direct calculation of (\ref{pwsterm}) gives the following source term
\begin{equation}\label{newsource}
\tilde S_l(k,r)=\frac{e^{ikr+\gamma\pi/2}}{2r^2\Gamma(-i\gamma)}.
\end{equation}
Interestingly the source term is independent from $l$. The derivation of  (\ref{newsource}) 
is given in Appendix \ref{appa}.

\section{Partial wave expansion and asymptotic forms}\label{cmpw}
The p.w. expansion of $\psi_{i}(\bk,\br)$ is given in \cite{har2} and we have reviewed it 
earlier. We now determine 
the corresponding expansion of the CMPW.
The p.w.  expansion of the CMPW is written in the standard form
\begin{equation}\label{taudef}
e^{i\bk\br}(kr -\bk\br)^{i\gamma}=\sum_{l=0}^\infty (2l+1)\tau_l(k,r)P_l(\cos(\vartheta)).
\end{equation}
and the radial functions are given by the integral
\begin{equation}\label{tau}
\tau_l(k,r)=\frac{1}{2}(kr)^{i\gamma}\int_{-1}^1e^{ikrx} (1- x)^{i\gamma} P_l(x) dx.
\end{equation}
A compact expression for the p.w. component of the CMPW can be given for arbitrary $l$. 
Using (\ref{tau}) and the integral 2.17.5.6 in \cite{prud} we get 
\begin{equation}\label{taunew}
\tau_l(k,r)=\frac{(-i\gamma)_l}{(1+i\gamma)_{l+1}}
(2kr)^{i\gamma}e^{ikr}\ _2F_2(1+i\gamma,1+i\gamma;l+2+i\gamma,1+i\gamma-l;-2ikr),
\end{equation}
where $(a)_n$ is the Pochhammer symbol. 

For the application of the complex scaling we have to know the asymptotic behavior 
of the scattered part of the 
wave function $\psi_l^{sc+}(k,r)$. Here we derive formulas valid at large $r$ values.
With the help of the expression  13.5.2 \cite{abra} we get the following asymptotic expansions valid at  $r\rightarrow\infty$
\begin{equation}\label{asycoui}
\omega_{i,l}(k,r)\sim\frac{e^{-ikr}(2kr)^{i\gamma}}{2ikr}(-1)^{l+1}
\sum_{n=0}^\infty a_{i,n}^l
\frac{1}{(2ikr)^n}
\end{equation}
and
\begin{equation}\label{asycous}
\omega_{s,l}(k,r)\sim\frac{e^{ikr}(2kr)^{-i\gamma}}{2ikr} e^{2i\sigma_l}
\sum_{n=0}^\infty a_{s,n}^l
\frac{1}{(2ikr)^n}.
\end{equation}
The expansion coefficients are given by $a_{i,n}^l=(-1)^n(l+1-i\gamma)_n(-l-i\gamma)_n/n!$ and 
$a_{s,n}^l=(l+1+i\gamma)_n(-l+i\gamma)_n/n!$.
These asymptotic forms show that 
with the help of the splitting (\ref{pwcousplit}) the incoming and outgoing spherical waves are clearly separated 
in the p.w. Coulomb-scattering wave function.

In order to derive an asymptotic expansion of $\tau_l(k,r)$ we express (\ref{taunew}) in terms of Meijer's $G$
function. Using 5.11.1(2) \cite{luke} we get
\begin{equation}
\tau_l(k,r)=e^{ikr}(2kr)^{i\gamma}(-1)^l
G_{2,3}^{1,2}\left(2ikr \left| 
\begin{array}{ccc}
-i\gamma, & -i\gamma\\
0, & -1-i\gamma-l, & l-i\gamma
\end{array}
\right.\right).
\end{equation}
We are interested in the asymptotic behavior after the CS is carried out i.e. $r$ is replaced by $re^{i\theta}$
and $0<\theta<\pi$. We give the asymptotic expansion valid in this case.
Considering the expression 6.5.32 \cite{abra} we can derive 
\begin{equation}\label{asytau}
\tau_{l}(k,re^{i\theta})\sim\frac{e^{-ikre^{i\theta}}}{2ikre^{i\theta}}(2kre^{i\theta})^{i\gamma}(-1)^{l+1}
\sum_{n=0}^\infty \frac{d^l_{n}(-2)^n}{(2ikre^{i\theta})^n},
\end{equation}
where the expansion coefficients satisfy the recursion
\begin{equation}
4(n+1)d^l_{n+1}=2[2n^2-n(2i\gamma-1)-l^2-l-i\gamma]d^l_{n}-n(n-l-i\gamma-1)(n+l-i\gamma)d^l_{n-1}
\end{equation}
and $d^l_{0}=1$.

\begin{figure}[tbp]
  \centerline{\includegraphics[scale=0.4]{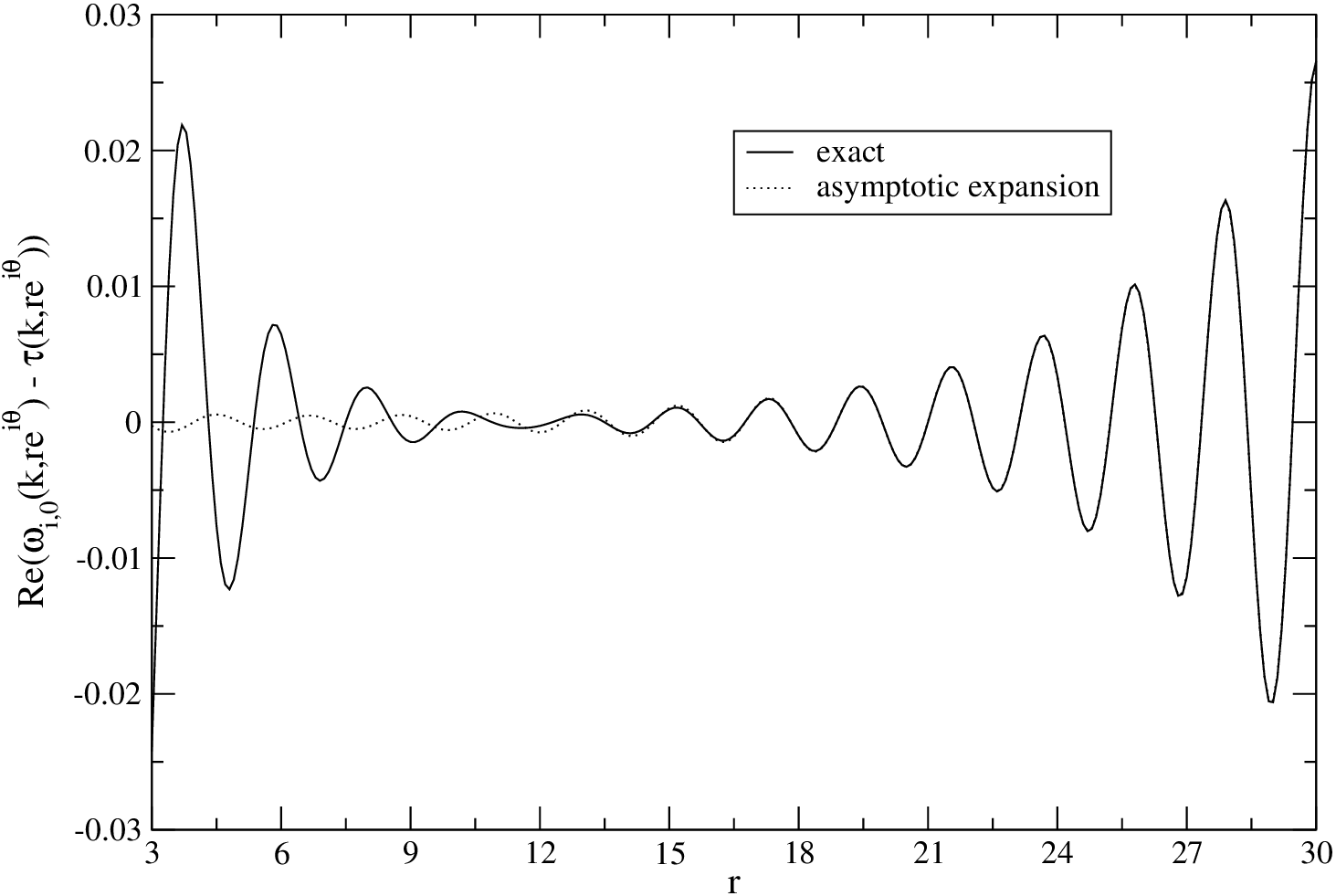}}
  \caption{The function $\omega_{i,0}(k,r)-\tau_0(k,r)$ and the next to leading order term of its asymptotic
  expansion (\ref{scatasy1}) are complex scaled using $\theta=0.1$. Only the real parts are displayed. 
  The momentum is $k=3$, the Sommerfeld-parameter $\gamma=1/3$ and $l=0$. The solid line denotes the exact values and the dashed
  line corresponds to the asymptotic expansion (\ref{scatasy1}).} 
\label{taufig}
\end{figure}

Since the p.w. components are related to each other by the 
simple relation
\begin{equation}\label{radparts}
\psi_l^+(k,r)=\tau_l(k,r)+\psi_l^{sc+}(k,r)
\end{equation}
and we have the splitting (\ref{pwcousplit}) we can write
\begin{equation}\label{scsplit}
\psi_l^{sc+}(k,r)=\left[\omega_{i,l}(k,r)-\tau_{l}(k,r)\right]+
\omega_{s,l}(k,r).
\end{equation}
We notice that the last term in (\ref{scsplit}) $\omega_{s,l}(k,r)$ asymptotically contains only 
outgoing spherical wave (see Eq. (\ref{asycous})) so the applicability of the CS is determined by the behavior of 
$\omega_{i,l}(k,re^{i\theta}))-\tau_{l}(k,re^{i\theta}))$ at $r\rightarrow\infty$.
Fortunately the asymptotic expansions of the functions $\omega_{i,l}(k,r)$ and $\tau_{l}(k,r)$ 
are carried out 
using the same asymptotic
sequence of functions $\{e^{-ikr}(2kr)^{i\gamma}(-1)^{l+1}/(2ikr)^n,\ n=0,1,2\ldots\}$ so we can simply add/subtract 
the asymptotic expansions as required \cite{erdelyi}.
Using (\ref{asycoui}) and (\ref{asytau}) we can write down the following asymptotic expansion 
\begin{equation}\label{scatasy1}
\left[\omega_{i,l}(k,re^{i\theta})-\tau_{l}(k,re^{i\theta})\right]\sim\frac{e^{-ikre^{i\theta}}}{2ikre^{i\theta}}
(2kre^{i\theta})^{i\gamma}(-1)^{l+1}\sum_{n=0}^\infty 
\frac{a_{i,n}^l-d^l_{n}(-2)^n}{(2ikre^{i\theta})^n}.
\end{equation}

Let's investigate  (\ref{scatasy1}). We realize that $a_{i,0}^l-d^l_{0}=0$. This means that 
in leading order $\psi_l^{sc+}(k,re^{i\theta})$ does not contain complex scaled incoming spherical wave. However in higher orders 
$\psi_l^{sc+}(k,re^{i\theta})$ do contains complex scaled "generalized" incoming spherical wave ($e^{-ikr}/r^n$, $n>1$). This means that 
the complex scaled scattered part of the wave function  $\psi_l^{sc+}(k,re^{i\theta})$ does not tend zero as 
$r\rightarrow\infty$. 

This finding is demonstrated in Fig. \ref{taufig}. 
Both the left hand side and the right hand side of 
(\ref{scatasy1}) are displayed. From the asymptotic expansion only the next to leading order term 
is considered (the leading order term is zero). The real part of the function
$\omega_{i,0}(k,re^{i\theta})-\tau_0(k,re^{i\theta})$ first starts to oscillate with decreasing order of amplitude 
however at larger $r$ values the presence of the terms of the form
$e^{-ikre^{i\theta}}/(re^{i\theta})^n$ dominate and the amplitude of the oscillation becomes larger and larger.

We have got a very unfortunate result, if we use the 3D splitting based on the CMPW then the 
scattered part of the wave function asymptotically contains both incoming and 
outgoing spherical waves. This fact prevents the application of the complex scaling.

\section{Complex scaling and scattering states}\label{cs}

In the previous section  we have established that the splitting of 
the wave function based on the CMPW i.e. the choice (\ref{cmpwsplit}) is useless  
from the point of view of CS. Now we turn to the splitting  (\ref{pwsumhar}) which is carried out on the 
p.w. level. The scattered part of the wave function is given by 
\begin{equation}\label{splituj}
\tilde\psi_l^{sc+}(k,r)=\omega_{s,l}(k,r)-\chi_l(k,r).
\end{equation}
This equation follows from (\ref{coui}), (\ref{pwcousplit}) and (\ref{pwsumhar}). 
The asymptotic form (\ref{asycous}) and the expression (\ref{pwchi}) for $\chi_l(k,r)$ shows that the scattered part 
of the wave function 
now contains only outgoing spherical wave and so the complex scaling can be safely applied. From (\ref{pwchi}) and 
(\ref{asycous}) we get in leading order
\begin{eqnarray}\label{sccsasy}
\tilde\psi_l^{sc+}(k,r) &=& e^{2i\sigma_l}\frac{e^{ikr}(2kr)^{-i\gamma}}{2ikr}
\left[ 1+O\left (\frac{1}{|2ikr|}\right)\right]\nonumber\\
&&-\frac{e^{ikr+\gamma\pi/2}}{2ikr}\frac{(-1)^l\Gamma(2l+1)}{\Gamma(l+1-i\gamma)}
\left[ \frac{(i\gamma-l)_l}{\Gamma(2l+1)}+O\left (\frac{1}{|2ikr|}\right)\right].
\end{eqnarray}

Let's make a variable transformation and replace $r$ with $re^{i\theta}$ in the partial wave driven Scr\"odinger equation 
(\ref{pwdriven}) and furthermore introduce a new function with the definition 
\begin{equation}\label{cswf}
\tilde\psi_{l,\theta}^{sc+}(k,r)=e^{i3\theta /2}\tilde\psi_l^{sc+}(k,re^{i\theta}),
\end{equation}
where $\theta$ is an arbitrary fixed real number. 
A simple calculation gives the following equation 
\begin{equation}\label{csdriven}
\left(\frac{k^2}{2}+e^{-2i\theta}\frac{1}{2r}\frac{d^2 }{dr^2}r-e^{-2i\theta}\frac{l(l+1)}{2r^2}
-e^{-i\theta}\frac{\gamma k}{r}\right)\tilde\psi_{l,\theta}^{sc+}(k,r)
= \tilde S_{l,\theta}(k,r),
\end{equation}
where the complex-scaled source term is defined by
\begin{equation}
\tilde S_{l,\theta}(k,r)=e^{i3\theta/2}\tilde S_l(k,re^{i\theta}).
\end{equation}
The advantage of the complex-scaled driven Scr\"odinger equation (\ref{csdriven}) is that its solution 
behaves very simply asymptotically. If the scaling angle satisfies the condition $0<\theta<\pi$ 
then from (\ref{sccsasy}) it follows  
\begin{equation}
\lim_{r \rightarrow \infty}\tilde \psi_{l,\theta}^{sc+}(k,r)=0.
\end{equation}
From the asymptotic form (\ref{sccsasy}) we can establish the following local representation of the 
partial wave Coulomb S-matrix
\begin{equation}\label{locrep}
e^{2i\sigma_l}\approx(2kre^{i\theta})^{i\gamma}\left[
e^{-ikre^{i\theta}}2ikre^{-i\theta/2}\tilde \psi^{sc+}_{l,\theta}(k,r)+
\frac{e^{\gamma\pi/2}(-1)^l(i\gamma-l)_l}{\Gamma(l+1-i\gamma)}\right]\ \ r \rightarrow \infty.
\end{equation}
The local representation of the phase shift given in \cite{Vol09} is different from (\ref{locrep})
since the splittings of the scattering wave function are distinct.

The function $\tilde\psi_l^{sc+}(k,r)$ is not regular at $r=0$. However the validity of the limit 
\begin{equation}\label{limitr0}
\lim_{r \rightarrow 0} r^{l+1}\tilde\psi_l^{sc+}(k,r)=0
\end{equation}
can be easily demonstrated. Details are given in Appendix \ref{appb}. 
In order to give simple boundary condition at $r=0$ we make the following transformation 
$h_{l,\theta}^{sc+}(k,r)=r^{l+1}\tilde\psi_{l,\theta}^{sc+}(k,r)$. This transformation 
leads to regular function at $r=0$.  For the new function we get the following differential equation
\begin{equation}\label{csdrivennew}
\left(\frac{k^2}{2}+e^{-2i\theta}\frac{1}{2}\frac{d^2 }{dr^2}
-e^{-2i\theta}\frac{l}{r}\frac{d }{dr}-e^{-i\theta}\frac{\gamma k}{r}\right)h_{l,\theta}^{sc+}(k,r)
= r^{l+1}\tilde S_{l,\theta}(k,r),
\end{equation}
The price we pay for the simplification at $r=0$ is the appearance of first order derivative in the equation.

From the earlier considerations presented it is obvious that the method based on the 
splitting (\ref{pwsumhar}) can be extended to case
when a short range interaction $V_s(r)$ is added to the pure Coulomb interaction. In
this case the inhomogeneous differential equation (\ref{csdrivennew})
is replaced by 
\begin{equation}\label{csdriventot}
\left(\frac{k^2}{2}+e^{-2i\theta}\frac{1}{2}\frac{d^2 }{dr^2}
-e^{-2i\theta}\frac{l}{r}\frac{d }{dr}-e^{-i\theta}\frac{\gamma k}{r}-V_s(re^{i\theta})\right)h_{l,\theta}^{sc+}(k,r)
= r^{l+1}S^{tot}_{l,\theta}(k,r),
\end{equation}
where the new source term reads
\begin{equation}
S^{tot}_{l,\theta}(k,r)=\tilde S_{l,\theta}(k,r)+e^{i3\theta/2}\psi_{i,l}(k,re^{i\theta})V_s(re^{i\theta}).
\end{equation}

\section{Numerical results}\label{numeric}

\begin{figure}[tbp]
  \centerline{\includegraphics[scale=.7]{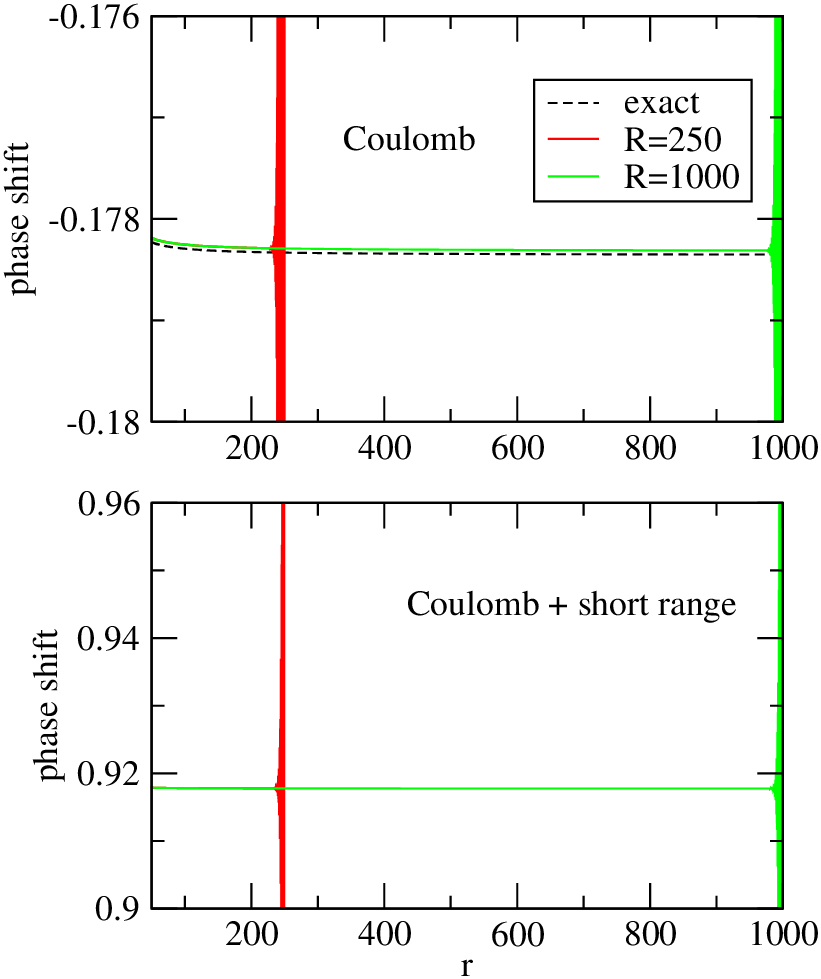}}
  \caption{The local representation of the phase shift for pure Coulomb potential (upper part) and  
 for the general case when a short range potential is added to the Coulomb term (lower part). In the first case the exact solution
 (dashed black line) is also displayed.
For the  numerical solution the boundary condition is imposed at R=250 (red line) and R=1000 (green line).
Detailed discussion is given in the text.
} 
\label{locrepfig}
\end{figure}

The differential equations (\ref{csdrivennew}) or (\ref{csdriventot}) have to be solved with the boundary conditions 
\begin{equation}\label{bcond1}
h_{l,\theta}^{sc+}(k,0)=0
\end{equation}
and
\begin{equation}\label{bcond2}
\lim_{r\rightarrow\infty}h_{l,\theta}^{sc+}(k,r)=0.
\end{equation}
In numerical calculations instead of (\ref{bcond2}) the boundary condition
\begin{equation}\label{bcond3}
h_{l,\theta}^{sc+}(k,R)=0.
\end{equation}
can be used. Here $R$ is a positive and otherwise arbitrary large number.
The boundary condition (\ref{bcond3}) is of course an approximation and it has  to be investigated how 
the result depends on $R$. The value of $R$
should be in the asymptotic region where (\ref{sccsasy}) is satisfied.

The finite element method is chosen as a numerical technique for the solution of Eqs. (\ref{csdrivennew}) or (\ref{csdriventot}).
The method and the basis functions used in any elements are described in \cite{res00}. The same method was used also in \cite{Vol09}.
For the presented calculations equally spaced finite elements of length 
1 is taken. The degree of the Lobatto shape functions \cite{res00} is denoted by $N$ and the same $N$ value is used at each elements. 
The $\theta$ parameter of the CS was chosen to 0.1 radian.

First the pure Coulomb case is considered i.e. the potential is given by $1/r$. In this case 
the numerical result can be compared to the known analytical solution. The momentum was $k=3$ and the considered orbital
angular momentum was $l=0$. The phase shift is calculated with the help of the local representation (\ref{locrep}).  
In this equation for $\tilde \psi^{sc+}_{l,\theta}(k,r)$  either the exact solution or the approximate one  
determined by the finite element method can be used.
In the second example the potential 
$7.5r^2\exp(-r)$ is added to the previous pure Coulomb term. Exactly these two cases were studied in \cite{Vol09} where a different splitting of 
the wave function and 
the exterior complex scaling method was used. 

\begin{figure}
  \centerline{\includegraphics[scale=0.4]{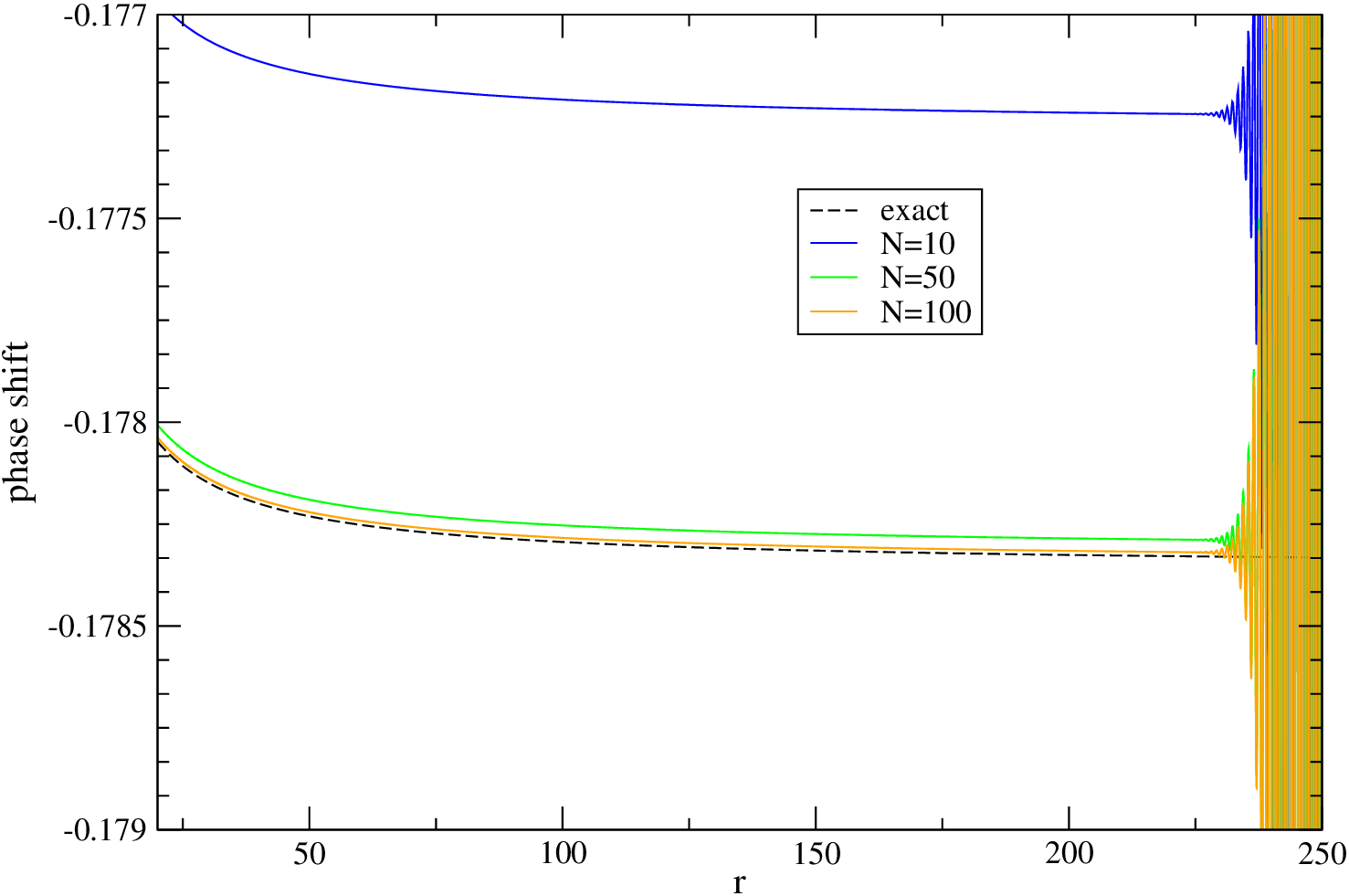}}
  \caption{The local representation of the phase shift for pure Coulomb potential.
In the  numerical solution the boundary condition is imposed at R=250. The exact solution and the numerical ones are displayed.
In the numerical calculations the number of the Lobatto shape functions $(N)$ is varied.
} 
\label{lobfig}
\end{figure}

The upper part of Fig. \ref{locrepfig} shows the results of the calculations carried out using 
pure Coulomb potential. 
In this case the exact solution (dashed black line) can be compared to the numerical ones.
In the finite element method the boundary condition (\ref{bcond3}) is imposed at two different $R$ values ($R=250$ and $R=1000$). We note that 
for $r>R$ the finite element solution is not defined.
The boundary condition should be set at infinity (see (\ref{bcond2})) but a finite  $R$ value is taken 
so it can be expected that the numerical solution 
is not accurate enough around the point where the boundary condition is set up. 
This can be clearly noticed in Fig. \ref{locrepfig}. If we choose  $R=250$  then 
there is an oscillation with large amplitude around $r=250$. 
If the boundary condition is set up at a larger $R$ value then
the oscillatory region is pushed out around this value. In  Fig. \ref{locrepfig}
the oscillatory region moved from $r=250$ to  $r=1000$ simply changing the value of the parameter 
$R$ from $R=250$ to $R=1000$. 
The effect of the boundary condition is noticeable. However, if this edge effect is not considered then 
the local representation of the phase shift is 
practically constant on a huge region. This is 
a useful feature since it helps to determine a unique value of phase shift of the numerical calculation.
In contrast the local approximation of the phase shift in \cite{Vol09} 
tends to the exact value by a persistent 
oscillation with decreasing order of amplitude. 

The lower part of  Fig. \ref{locrepfig} shows the results when 
the short range potential is added to the Coulomb interaction.
This modification 
does not change the previous observations.
The lower part of Fig. \ref{locrepfig} clearly demonstrates each previous conclusions. 
The position of the boundary condition influences the value of the phase shift
only around the point $r=R$. The phase shift calculated by the expression (\ref{locrep}) is practically independent from the value of $r$. 
We note that 
apart from the oscillatory region around $r=250$ the two numerical solutions corresponding to 
the choices $R=250$ and $R=1000$   
coincide in the region $50<r<250$.  In this region in Fig. \ref{locrepfig} the red and green lines are 
indistinguishable on the used scale.

The calculations displayed in Fig. \ref{locrepfig} have been carried out using 50 Lobatto shape 
functions on each elements. We investigated the 
dependence of the local representation of the phase shift on the number of Lobatto functions used 
in the finite element method.
The boundary condition (\ref{bcond3}) is set up at $R=250$. The results are depicted in Fig. \ref{lobfig}. 
Apart from the region around $r=250$ the exact phase shift is 
reproduced with three digits accuracy with $N=50$. 
The calculation with $N=100$ 
very well reproduces the exact solution (4 digits agreement is reached almost everywhere).

\begin{figure}
  \centerline{\includegraphics[scale=0.4]{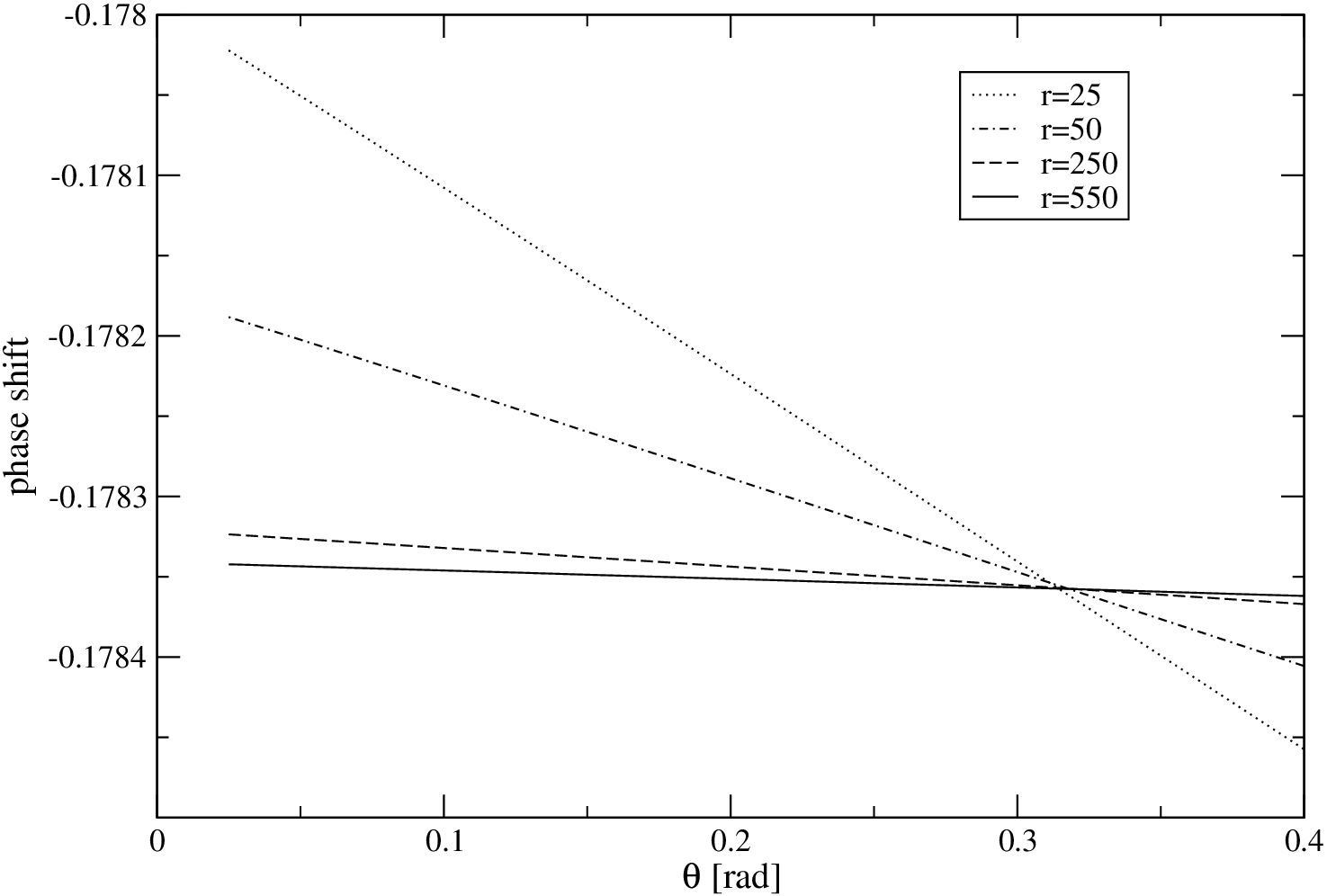}}
  \caption{The local representation of the phase shift for pure Coulomb potential as the function 
  of the complex scaling parameter. Four different $r$ values are considered and in the local 
  representation of the phase shift the exact wave 
  function is used.
} 
\label{thetafig}
\end{figure}

It remains to check how the complex scaling parameter $\theta$ influences the calculated phase shift. 
Figure \ref{thetafig} displays 
the local representation of the phase shift as the function of the complex scaling parameter. Four different $r$ values are used.
In these calculation the exact wave function is used in (\ref{locrep}). 
If the value of $r$ is  
in the asymptotic region (e.g. $r=550$) then 
the calculated phase shift is
independent from the value of the complex scaling parameter. For smaller $r$ values it is advantageous to use larger $\theta$ value to get 
better agreement with the exact phase shift. 

\section{Summary}\label{sum}
We have rigorously shown that the two-body scattering problem of the pure Coulomb interaction 
can be solved using the standard complex scaling method. This is achieved without using any
cutoff of the long range interaction.
The intricate scattering boundary condition is greatly simplified and so 
the numerical solution can be plainly achieved.
It is obvious that the suggested  driven Scr\"odinger equation 
can be solved by the use of the
exterior complex scaling method too.
The advocated splitting of the total
wave function works for general circumstances. It can be applied not only for pure Coulomb
force but short range interactions can be added to the Coulomb potential. It turned out that the
splitting based on the Coulomb modified plane wave does not lead to simplification of the boundary
condition from the point of view  of the complex scaling. 

\appendix
\section{}\label{appa}
In the case of the splitting (\ref{pwsumhar}) according to (\ref{pwsterm}) the source term
reads
\begin{equation}\label{pwstermc1}
\tilde S_l(k,r)=\left[-\frac{1}{2r}\frac{d^2 }{dr^2}r
+\frac{l(l+1)}{2r^2}+\frac{\gamma k}{r}-\frac{k^2}{2}\right][\omega_{i,l}(k,r)+\chi_l(k,r)].
\end{equation}  
Since  $\omega_{i,l}(k,r)$ is just the Coulomb Jost solution \cite{har,har2} the 
contribution from $\omega_{i,l}(k,r)$ to the source is zero. Introducing a new variable $z=2ikr$ and 
rewriting the summation in (\ref{pwchi}) we have
\begin{equation}\label{pwstermc2}
\tilde S_l(k,r)=\left[-\frac{k^2}{2}+\frac{2k^2}{z}\frac{d^2 }{dz^2}z
-\frac{2k^2l(l+1)}{z^2}+\frac{2i\gamma k^2}{z}\right]\chi_l(z),
\end{equation}  
where
\begin{equation}
\chi_l(z)=\frac{e^{\gamma\pi/2}}{\Gamma(1-i\gamma)}\frac{e^{z/2}}{z}\sum_{n=0}^l
\frac{(l+1)_n(-l)_n(1)_n}{(1-i\gamma)_n}\frac{z^{-n}}{n!}.
\end{equation}
Direct calculation gives
\begin{equation}
\tilde S_l(k,r)=\frac{2k^2e^{\gamma\pi/2}}{\Gamma(1-i\gamma)}\frac{e^{z/2}}{z^2}\sum_{n=0}^l
\frac{(l+1)_n(-l)_n}{(1-i\gamma)_n}
\left(i\gamma-n+
\frac{n(n+1)-l(l+1)}{z}\right)z^{-n}.
\end{equation}
Rearranging the summation we get 
\begin{eqnarray}
\tilde S_l(k,r) &=& \frac{2k^2e^{\gamma\pi/2}}{\Gamma(1-i\gamma)}\frac{e^{z/2}}{z^2}\times \nonumber\\
&&\qquad\times\left\{i\gamma+\sum_{n=0}^{l-1}\left[\frac{(l+1)_n(-l)_n}{(1-i\gamma)_n}[n(n+1)-l(l+1)]+\right.\right.\nonumber\\
&&\left.\left.\qquad\qquad\qquad\qquad\qquad+(i\gamma-n-1)\frac{(l+1)_{n+1}(-l)_{n+1}}{(1-i\gamma)_{n+1}}
\right]z^{-n-1}\right\}.
\end{eqnarray}
Using the fact that the Pochhammer symbols satisfy the recursion $(a)_{n+1}=(a)_n(a+n)$ we can show that 
the expression inside the square bracket is zero and so we have proved (\ref{newsource}).

\section{}\label{appb}

According to (\ref{cous}) in order to prove (\ref{limitr0}) it is enough to show that
\begin{equation}
\lim_{r\rightarrow 0}r^{l+1}\omega_{s,l}(k,r)=\lim_{r\rightarrow 0}r^{l+1}\chi_{l}(k,r).
\end{equation}
From the definition (\ref{pwchi}) it follows 
\begin{equation}\label{lim1}
\lim_{r\rightarrow 0}r^{l+1}\chi_{l}(k,r)=e^{\gamma\pi/2}\frac{(-1)^l}{(2ik)^{l+1}}
\frac{\Gamma(2l+1)}{\Gamma(l+1-i\gamma)}.
\end{equation}
Considering (\ref{pwcous}) and the expression 7.2.2.3 \cite{prud} the function $r^{l+1}\omega_{s,l}(k,r)$ is a sum of
three terms
\begin{equation}
r^{l+1}\omega_{s,l}(k,r)=A_l(k,r)+B_l(k,r)+C_l(k,r),
\end{equation}
where
\begin{equation}
A_l(k,r)=-K_l(k)\frac{M(l+1+i\gamma,2l+2;-2ikr)}{\Gamma(2l+2)\Gamma(i\gamma-l)}(-2ikr)^{2l+1}\log(-2ikr)e^{ikr},
\end{equation}
\begin{equation}
B_l(k,r)=-K_l(k)\frac{e^{ikr}}{\Gamma(2l+2)\Gamma(i\gamma-l)}\sum_{n=0}^\infty f^l_n\frac{(l+1+i\gamma)_n}{(2l+2)_n}
\frac{(-2ikr)^{n+2l+1}}{n!}
\end{equation}
and
\begin{equation}
C_l(k,r)=-K_l(k)\frac{(2l)!e^{ikr}}{\Gamma(l+1+i\gamma)}\sum_{n=0}^\infty \frac{(i\gamma-l)_n}{(-2l)_n}
\frac{(-2ikr)^n}{n!}.
\end{equation}
The following abbreviations are used
\begin{equation}
K_l(k)=\frac{(-1)^{l+1}}{(2ik)^{l+1}}e^{2i\sigma_l+\gamma\pi/2}
\end{equation}
and
\begin{equation}
f^l_n=\Psi(l+1+i\gamma+n)-\Psi(n+1)-\Psi(2l+2+n).
\end{equation}
The digamma function is denoted by $\Psi(z)$. From the expressions above it is clear that 
$\lim_{r\rightarrow 0}A_l(k,r)=0$ and $\lim_{r\rightarrow 0}B_l(k,r)=0$. The limit value of $C_l(k,r)$
as $r\rightarrow 0$ is exactly the right hand side of (\ref{lim1}) so we have proved  (\ref{limitr0}).

\begin{acknowledgments}

This research was supported by Hungarian OTKA Grant No. T46791.

\end{acknowledgments}

\end{document}